# Fair and Efficient Ridesharing: A Dynamic Programming-based Relocation Approach

AQSA ASHRAF MAKHDOMI, NIT Srinagar, India
IQRA ALTAF GILLANI, NIT Srinagar, India

Recommending routes by their probability of having a rider has long been the goal of conventional route recommendation systems. While this maximizes the platform-specific criteria of efficiency, it results in sub-optimal outcomes with the disparity among the income of drivers who work for similar time frames. Pioneer studies on fairness in ridesharing platforms have focused on algorithms that match drivers and riders. However, these studies do not consider the time schedules of different riders sharing a ride in the ridesharing mode. To overcome this shortcoming, we present the first route recommendation system for ridesharing networks that explicitly considers fairness as an evaluation criterion. In particular, we design a routing mechanism that reduces the inequality among drivers and provides them with routes that have a similar probability of finding riders over a period of time. However, while optimizing fairness the efficiency of the platform should not be affected as both of these goals are important for the long-term sustainability of the system. In order to jointly optimize fairness and efficiency we consider repositioning drivers with low income to the areas that have a higher probability of finding riders in future. While applying driver repositioning, we design a future-aware policy and allocate the areas to the drivers considering the destination of requests in the corresponding area. Extensive simulations on real-world datasets of Washington DC and New York demonstrate superior performance by our proposed system in comparison to the existing baselines.

## 1 INTRODUCTION

The rapid proliferation of mobile service providers has had a significant impact on the way people connect with their surroundings, and one of the prime examples of this shift is the rise of ride-hailing platforms. These services have disrupted the traditional taxi industry by providing a convenient and efficient way for people to get around. The ability to request a ride through a mobile app and track the driver's location in real-time has made the process of hailing a ride seamless. Urbanization and the emergence of convenient fee structures have further led to an increase in the use of these services for daily commutes, which has resulted in the shortage of vehicles during high-demand hours. To address the problem of increasing demand and utilize the vehicle in an effective manner, ridesharing has been proposed as a solution that pairs multiple riders with similar time itineraries in a single vehicle, which not only addresses the shortage of vehicles during peak hours but also reduces the carbon footprints by decreasing the count of vehicles on the road [11]. Moreover, ridesharing services reduce the fare in comparison to their counterpart solo rides, which makes them a popular choice among riders.

Despite their popularity, there have been various reports about discrimination against drivers in well-known transport network companies like Uber and Ola [13]. It has been found that there is a vast difference in the earnings of drivers working for similar time periods [17, 30]. The root cause of this discrimination is the profit-driven model employed by these companies, which prioritizes matching drivers with riders in a quick manner in order to enhance customer satisfaction and

Authors' addresses: Aqsa Ashraf Makhdomi, makhdoomiaqsa@gmail.com, NIT Srinagar, Hazratbal, Srinagar, Jammu & Kashmir, India, 190020; Iqra Altaf Gillani, iqraaltaf@nitsri.ac.in, NIT Srinagar, Hazratbal, Srinagar, Jammu & Kashmir, India, 190020.

ultimately increase the customer base. This results in discrimination among drivers based on their locations, where the drivers operating in areas dense in requests receive numerous passenger requests, and on their counterpart, the drivers in sub-urban areas fail to secure rides. It creates income differences among them where the income of drivers depends upon their probability of being at an area dense in requests, and not on their working hours. This form of discrimination against drivers undermines the basic principle of fair competition and can lead to a lack of trust in the platform. Fairness is characterized by the impartial treatment of individuals or groups, irrespective of their inherent or acquired characteristics like their location. Our proposed model ensures a fair route recommendation system is developed for drivers where their earnings are proportional to the number of hours they work and not on their probability of being at a high-demand location.

There has been growing research on effective route recommendation strategies [5, 16, 28] in ride-hailing platforms. However, these works are designed to increase the customer base of the platform and have revolved around the parameters that bring more profit [16, 28] to the drivers or reduce the waiting time of passengers [5]. Although there are some works that have designed fair route recommendation systems for ride-hailing platforms [15, 17], these algorithms don't take into account the constraints of riders time schedules and their distances [12]. We propose the design of a route recommendation system for ridesharing platforms that considers the time schedules of different riders sharing a ride and returns a set of routes that are fair towards the drivers.

Fairness is an important criterion and should be satisfied for the long-term sustainability of the system. However, while fairness is important, it is not the only goal of the system. Efficiency is also a critical consideration, and both these objectives must be jointly optimized for the system to function optimally. We propose to optimize these objectives, by designing a relocation and priority-based framework. The priority-based approach systematically addresses fairness concerns over time, fostering an environment where drivers with lower incomes have a high probability of receiving passenger requests compared to their higher-earning counterparts. The strategic prioritization of lower-earning drivers in recommendations serves a dual purpose. Firstly, it rectifies the existing disparity by providing a fair share of opportunities to those with lower incomes. Secondly, it optimizes platform efficiency by providing low-earning drivers with routes that have a higher flow of passengers, which results in an increase in service coverage of the platform. While the priority approach aims to grant low-earning drivers a higher recommendation priority, a potential drawback arises when these drivers find themselves in low-demand areas. In such instances, the priority mechanism may recommend optimal routes within those areas, leading to prolonged inefficiencies. To counteract this issue, our proposed approach incorporates a relocation strategy. Here, drivers are strategically relocated to areas anticipated to experience high passenger demand. This dual approach optimizes fairness and efficiency by preventing the persistence of suboptimal routes in low-demand regions and effectively balancing demand and supply across different areas.

As the problem of route recommendation is NP-Hard [11], no polynomial time algorithm exists that can solve the problem on a deterministic machine. We propose to improve the complexity of the recommendation by constraining the search space to a Directed Acyclic Graph (DAG) and applying dynamic programming to it. Our main contributions can be summarized as follows:

- To the best of our knowledge, this is the first work that considers the design of a *fair* route recommendation system for *ridesharing* networks.
- The underlying problem is NP-Hard and we propose to constrain the search space from the entire road network to a Directed Acyclic Graph (DAG). Thereafter, we apply dynamic programming to identify the optimal path on this simplified graph.
- The proposed route recommendation system jointly optimizes system efficiency and fairness by designing a future-aware relocation policy for drivers.

- We demonstrate the performance of our proposed model on the two real-world datasets from Washington DC and New York City. Experimental results show that our proposed model is able to improve the efficiency and fairness of the system by a significant margin as compared to the existing baselines.

## 2 RELATED WORK

Our research primarily pertains to the following fields of study: 1) Route recommendation for ride-hailing (solo) and ridesharing (pool) platforms and 2) Fairness in ride-hailing platforms.

### 2.1 Route recommendation for ride-hailing and ridesharing platforms

In recent years, there has been a surge in research focused on the design of route recommendation systems for ride-hailing and ridesharing platforms. The objective of these studies is to either maximize the platform's profit or reduce the waiting time of riders. To achieve the above goals, these studies predict the future demand for rides and recommend routes with a higher probability of having rides on the way. In this direction, Garg *et al.* [5] proposed a route recommendation system for idle taxi drivers that aimed at minimizing the distance between drivers and anticipated riders. They employed a Monte Carlo search tree to predict future rider requests and proposed a shortest passenger-aware path, that resulted in increased profit for drivers and reduced the waiting times of riders. However, a significant drawback of their model is the high computational cost in dynamic environments which makes it impractical to implement. Qu *et al.* [16] proposed a profitable route recommendation system for ride-hailing platforms that allocated the routes based on the supply and demand at that point in time. The proposed model used a probabilistic approach to balance the drivers and riders and recommended the path to the driver based on the expected cruising distance from the rider. Additionally, the authors applied Map-Reduce and kdS-tree algorithms to improve the recommendation efficiency and overcome the computation complexity problems of the previous models. However, these models are designed for ride-hailing platforms which operate with a single rider in the vehicle.

Ridesharing has gained widespread popularity as a convenient and cost-effective transportation mechanism. Numerous studies have explored matching algorithms and route planning frameworks that take into account factors such as dynamic pricing [1], rider preferences for social comfort [3], or a combination of both [7, 9]. However, these models primarily focus on matching drivers with riders and do not recommend routes to drivers. Designing a route recommendation system for ridesharing platforms is a complex task that involves optimal utilization of the vehicle capacity within the constraints of the time schedules of different riders and the distance travelled by each of them. In this direction, Schreieck *et al.* [18] proposed a matching algorithm and a route recommendation system based on Dijkstra's algorithm. Their proposed model created a sustainable environment through the efficient utilization of vehicle. However, it required a significant amount of memory attributed to the larger size of nodes within the road network, which resulted in inefficiency over large-scale networks. Ta *et al.* [23] proposed a new ridesharing model that maximized the overall shared route percentage and alleviated traffic pressure. Their proposed method was computationally efficiently as it pruned the search space and eliminated the infeasible driver-rider pairs. However, their model is designed for matching drivers and riders and they do not provide route recommendation to drivers. Yuen *et al.* [28] proposed a route recommendation system for ridesharing networks that aimed to optimize the utilization of vehicles and reduce greenhouse emissions. Their proposed approach predicted the sequence of pick-up points that spanned a length greater than the shortest path and had an enhanced probability of picking up potential riders on the way. Tong *et al.* [25] proposed an insertion-based route planning framework that recommended a route based on the expected passenger demand and inserted a new request into the existing route through its insertion operation. However, the above studies [25, 28] have optimized the efficiency of the platform by predicting the passenger demand (origin of passengers) in different regions, whereas the passenger flow can be represented if the origin, as well as the destination of passengers, is known beforehand. We provide an efficient route recommendation system that predicts the origin and destination of requests using the GNN-based model described in [10] and applies dynamic programming to constrain the search space.

### 2.2 Fairness in ride-hailing platforms

The issue of fairness in the ride-hailing industry has been steadily gaining attention and causing increasing concern [30]. Studies have revealed a notable disparity in the income earned by drivers who work for comparable time periods [4, 30]. These concerns stem from the profit-driven models employed by the platform, which match the drivers with riders based on their location. Thus, if a driver is in an urban area his probability of getting matched to a rider is high as compared to his counterpart who is in a suburban area. This results in income disparity among drivers working for similar time periods. In order to overcome this algorithmic bias and ensure the drivers who work for the same number of hours get the same income irrespective of their location, some of the recent works [6, 17, 20, 22] have proposed models that match the drivers and riders keeping in perspective the fairness criterion. Lesmana *et al.* [6] introduced a re-assignment mechanism aimed at optimizing the balance between efficiency and fairness in the earnings of ridesharing platform drivers. Their approach initially matched the drivers and riders based on the utility of drivers, and if the matching values fell below a fairness threshold, the model facilitated fairer matches by implementing re-assignments. Rong *et al.* [17] developed a centralized taxi dispatching algorithm that aimed to optimize drivers' revenue while taking into account the fairness of the platform. To achieve this, they implemented a dynamic system that calculated the priority of drivers based on their income and matched them with riders according to their priority levels. Additionally, they designed a route recommendation system for the drivers that were unable to find a match with riders. Shi *et al.* [20] proposed a fair and efficient task assignment scheme that used Reinforcement learning to balance the income among a set of drivers. Their proposed model considered the future assignment of requests and employed various acceleration techniques to achieve fast and fair assignments on large-scale data. Sun *et al.* [22] introduced a multi-agent reinforcement learning framework to incorporate fairness in driver earnings within a ride-hailing platform. Their approach involved treating each driver as an agent and training a distributed policy for them in order to jointly optimize fairness and efficiency of the platform. However, these studies have focused on matching drivers with riders and have not considered the constraints on riders' time schedules. Apart from the fair matching algorithms proposed for ride-hailing platforms some of the recent works in spatial crowdsourcing [2, 8, 27, 29] have incorporated fairness and ensured the tasks allocated to drivers are fair and do not result in differences in income among them. However, while these models address income equality through task allocation and route planning, they do not offer mechanisms for recommending routes to the workers.

In the direction of fair route recommendation systems, limited work has been done. Qian *et al.* [15] introduced a fair route recommendation system for drivers of the ride-hailing platforms. Their approach assigned routes based on rider count and driver proximity while ensuring fairness over time. It prevented a single driver from consistently receiving the best routes, promoting overall fairness. However, the proposed approach is designed for ride-hailing platforms and does not consider the ridesharing environment when there are multiple riders in the vehicle with different time schedules. We consider the design of a fair and efficient route recommendation system for the drivers of ridesharing platforms that considers the constraints of rider schedules and reduces the income differences among them.

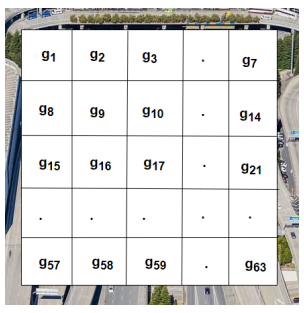

Fig. 1. Road network in the form of grid

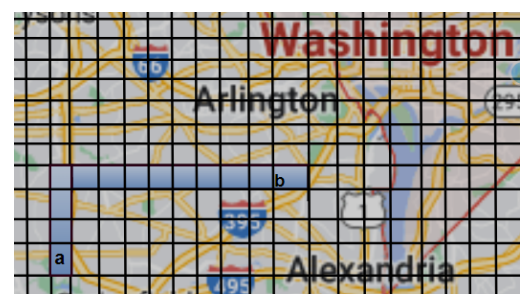

Fig. 2. Mapping of grid to a road network

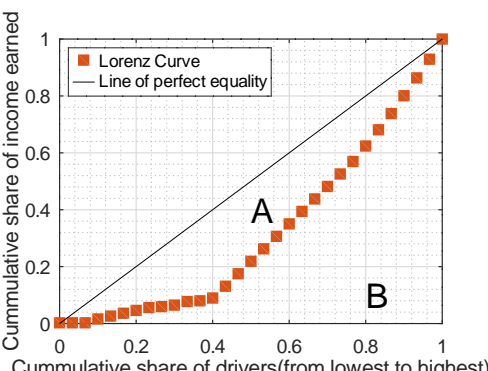

Fig. 3. Lorenz curve

## 3 PRELIMINARIES

In this section, we will introduce the concepts that are necessary for understanding the working of our proposed model.

**Grid**. As the road network is vast, it is modelled by dividing the entire area into a grid which is a collection of $m$ non-overlapping grid cells, represented as $g = \{g_1, g_2, ..., g_m\}$. These cells are said to be connected if they are located next to each other. For instance, consider the part of the road network represented in the form of a grid in Figure 1. The network is a combination of 63 grid cells and grid cell $g_1$ is connected to grid cells $g_2, g_8$ and $g_9$.

Prior research [22, 24–26] in the field of route recommendation systems predominantly employed a grid-based methodology to suggest optimal routes. This approach partitions the road network into discrete geographical zones and provides a simplified representation of the road infrastructure. On this network, drivers are directed towards particular zones marked by a high volume of ride requests. Though this approach does not provide exact routes to drivers, it recommends the areas with the higher flow of requests which helps drivers to reach out to the areas where demand will be high. Figure 2 shows the mapping of the Washington DC dataset into a grid. This figure displays how the entire area is covered by the grid cells. For instance, if the driver has to travel from grid cell $a$ to $b$ the sequence of grid cells that the driver follows is highlighted in blue.

When selecting a grid as a measure of recommending routes, the size of the grid cell plays an important role. If the size of the grid cell is large then multiple areas will be concentrated within a single grid cell and the recommendation mechanism will not be efficient. On the other hand, if the size of the grid cell is small, the recommendation system will be complex as the size of the network increases. Moreover, with the smaller grid cell size, the prediction accuracy of the underlying GNN-based prediction model [10] also decreases. Keeping these points in place, we have followed the directions of previous studies on route recommendation systems and matching algorithms [24, 26] and kept the grid cell size as 1.24 miles.

**Route**. A route corresponds to a path on the graph and it is a sequence of connected grid cells, represented as $P = \{g_1, g_2, ...g_l\}$, where $l \leq m$.

**Ride order and detour ratio** . A ride-order $o$ also referred to as a request is represented by a tuple $< o_s, o_d, o^t >$ where $o_s, o_d$ denote the source and destination of the request respectively and $o^t$ denotes the time at which the request is made.

When a route is recommended to the driver, it can't pick up all the ride orders on that route. This is because the destinations of different ride orders will be different and reaching the destination of each order will probably take a fraction of different route which will result in an increased distance travelled by the riders. In order to ensure the extra distance travelled by riders is bounded, their distance between source and destination points should be closest to the shortest path between those points. This is ensured by using the *detour ratio* ($\alpha$) which is the ratio of distance travelled by the rider between its origin and destination points to the shortest distance between the corresponding points. Mathematically, it is represented as $\alpha(P, o_s, o_d) = \frac{|P(o_s,o_d)|}{|SP(o_s,o_d)|}$. It is a function of the path $P$ taken by the driver between the vertices $o_s$ and $o_d$, and the shortest path between those vertices. To be specific, $|P(o_s, o_d)|$ denotes the distance travelled by the rider when travelling between vertices $o_s$ and $o_d$ through path $P$, and $|SP(o_s, o_d)|$ denotes the distance of the shortest path between the corresponding vertices.

When a new ride order arrives the path $P$ taken by the vehicle changes to incorporate the order. In order to ensure riders that are in the vehicle are satisfied, the increase in distance travelled through the modified path $P$, should be closest to the distance of the shortest path between their endpoints $o_s$ and $o_d$. This is ensured by setting the detour ratio of all ride orders in the vehicle ($o_s \in V$) below a threshold value $t_d$. It establishes that different ride-orders share routes and the increase in distance travelled by taking the new order is bounded. Mathematically it is represented as:

$$\forall_{o_s \in V} \ \alpha(P, o_s, o_d) \ \leq t_d \tag{1}$$

**Trip Fare.** The fare of a ride depends upon various factors, which include the distance travelled by the rider $j$, denoted as $d_j$, the fare charged per unit of distance, $f$, the duration for which the rider $j$ is in the vehicle, represented as $t_j$, and the fare charged per unit of time, $f'$. In addition to the above factors, there are two other values that are applied to a ride: a base fare $b_f$, which is determined by the distance between the pick-up location of the rider and the drivers current location, and a minimum fare $m_f$, that the rider must pay if the fare falls below its value. Mathematically, the fare is represented as:

$$f_j = \max(m_f, b_f + d_j \cdot f + t_j \cdot f') \tag{2}$$

**Utility.** The utility of a driver is a measure of his satisfaction with the platform, and we assume the drivers satisfaction depends upon his earnings. The primary factors that influence the earnings of driver include the number of rides $n$ taken by him, revenue $r_k$ generated from ride $k$, distance $d_k$ travelled when delivering riders in ride $k$, and the cost per unit of distance $c$. Mathematically, it is represented as:

$$u(r, d, c, n) = \sum_{k=1}^{n} (r_k - d_k \cdot c) \tag{3}$$

The revenue ($r_k$) generated by the driver from a ride $k$ depends upon the number of ride orders ($n_k$) served by the driver in a single ride and the fare of each order ($f_j$). If there is a single rider in the vehicle, then the revenue is simply the fare of that rider. Since we are enabling ridesharing there can be multiple riders in the vehicle. Consequently, the fare of a ride decreases if the rider gets paired with another rider on his way to the destination. We follow the ridesharing model by Uber [21] wherein the decrease in the fare of a ride is by 20% when the rider gets paired with another rider on his way to the destination. Mathematically, the revenue $r_k$ generated from a ride $k$ is defined as:

$$r_k = \begin{cases} \sum_{j=1}^{n_k} 0.8 \cdot f_j & if n_k > 1 \\ f_j & otherwise \end{cases} \tag{4}$$

In addition to the revenue generated, the driver will develop some expenses to cover the cost of fuel when delivering the riders to their destinations. These expenses are subtracted from the revenue and they include the distance $d_k$ travelled by the driver when delivering riders in ride $k$ and the cost per unit of distance $c$.

Next, we will discuss the concepts related to fairness.

**Lorenz curve**. Lorenz curve is a graphical representation that is used to display the distribution of wealth or income of a population. The curve plots the cumulative share of the total variable (such as income) against the cumulative share of the population, ranked by the same variable. This curve is used by our proposed system to display the level of income that is concentrated within the different fractions of drivers. It is depicted through Figure 3, where the X-axis represents the fraction of drivers, Y-axis represents the income, and the curve displays the income share within the hands of a fraction of the population. In a fair distribution also called the line of perfect equality, the Lorenz curve would form a straight line at a 45-degree angle, indicating the equal distribution of the variable among the population. However, in real-world distributions, the curve slopes downward, reflecting that a smaller portion of the population holds a disproportionate share of the income. The difference between the 45-degree line and the actual Lorenz curve represents the level of inequality in the distribution.

**Gini coefficient**. Gini coefficient serves as a metric for assessing the level of inequality in the distribution of income or wealth of a population. In our proposed system, we have used it to display the inequality in the income of drivers. A Gini coefficient of 0 represents complete equality in the distribution, while a value of 1 reflects the highest level of inequality. Mathematically, its value quantifies the degree of spread of the Lorenz curve from the ideal case (line of perfect equality). It is calculated as a ratio of two areas: the area between the Lorenz curve, and the line of perfect equality (denoted as A in Figure 3), to the area below the line of perfect equality ( $A+B$ in Figure 3).

## 4 EFFICIENT ROUTE RECOMMENDATION SYSTEM

In this section, we will describe the model that maximizes the efficiency of the system. We define the problem of efficiency maximization through the concepts from welfare economics and thereafter apply dynamic programming to reduce its complexity from the exponential frame to a polynomial size.

### 4.1 Problem Formulation

The efficiency of the proposed model is formulated by the Utilitarian criterion of social welfare economics, which is a philosophical approach that evaluates the moral worth of an action based on its ability to maximize the overall happiness or pleasure of people. It is one of the axiomatically justified units of efficiency and has been extensively studied in the literature. According to it, the platform's efficiency is the cumulative utility of the drivers, and the platform will be better off if it is maximized. Thus the objective function of the platform is framed as

$$max \sum_{i=1}^{n_d} u_i(r, d, c, n) \tag{5}$$

This criterion tries to maximize the sum of the utilities of all the $n_d$ drivers who are in the platform. To achieve this, the proposed model needs to recommend routes to a set of competing drivers that have a higher probability of finding riders on the way. This leads to the efficient utilization of vehicles and results in an increase in the drivers' profits. However, while optimizing this objective, the system should ensure rider satisfaction and recommended routes should not incur excessively long distances. In order to effectively model the passenger requests and distance constraints of ridesharing platforms, the proposed system requires two subgraphs-one for representing the passenger requests and the other for determining the distance travelled by the passengers. We will describe the creation of these graphs in detail next.

*4.1.1 Graphical Modelling.* The proposed framework is modelled through a family of subgraphs $G = \bigcup_{k \in \{Q,R\}} G^k = \bigcup_{k \in \{Q,R\}} (V, E^k, w^k)$, where the vertices of the subgraph $V$ represent the grid cells and the edges $E$ represent the connections that can arise between various grid cells. There are two subgraphs, one is the request graph $G^k = G^Q$ that is used to recommend a route with the highest number of *expected* passenger requests, and the other is the road graph $G^k = G^R$ which keeps a constraint on the distance travelled by each passenger. The edge weight $w_{ij}^Q$ between the vertex $v_i$ and $v_j$ of request graph $G^Q$ represents the *expected* number of requests that can originate between the specified source vertex ($v_i$) and destination vertex ($v_j$), and these requests are predicted through the GNN-based model described in [10]. In our proposed model, we have predicted the origin as well as the destination of requests and utilized it as an edge weight because the studies have found

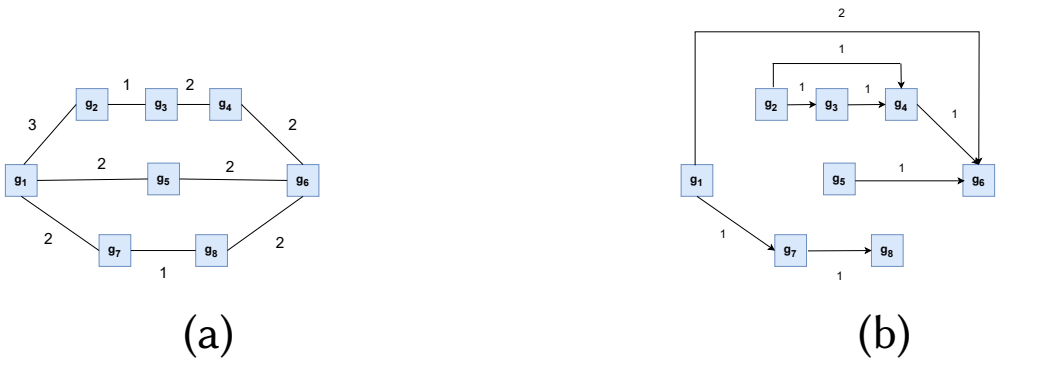

(a) (b)

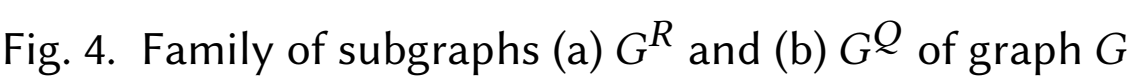

Fig. 4. Family of subgraphs (a) $G^R$ and (b) $G^Q$ of graph $G$

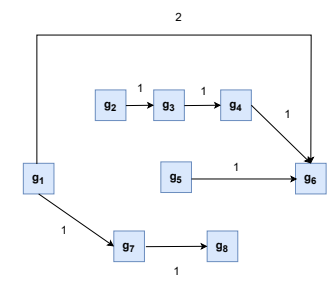

Fig. 5. Directed Acyclic Graph of request graph shown in Figure 4b

[11] that knowing the origin and destination of requests in advance can help in effective vehicle utilization in ridesharing platforms. For the road graph $G^R$, the edges represent the connections between adjacent grid cells and the edge weight determines the distance between them.

Figures 4a and 4b show an instance of road graph and the corresponding request graph. The vertices in both the subgraphs represent grid cells, but the edges and their weights are unique to each. In the subgraph $G^R$, the presence of an edge signifies a connection between two grid cells, with the weight of an edge indicating the distance between them. For instance, in the road instance displayed by Figure 4a, the distance between grid cells $g_2$ and $g_3$ is 1. The request graph $G^Q$, on the other hand, is a complete graph where requests can come between any two vertices. The edges in this graph are directed and the direction of an edge indicates the flow of requests. As an example, consider the request graph shown in Figure 4b. For clarity, we have ignored the edges with 0 weight. The edge between the grid cells $g_1$ and $g_6$ has a weight of 2, indicating that there are 2 requests with $g_1$ as their starting point and $g_6$ as their destination.

*4.1.2 Graph traversing.* In order to maximize the utilities of drivers, the proposed model needs to select a path from the request graph that has the highest number of expected requests. However, selecting a path with the maximum requests may lead to passenger dissatisfaction as they may have to travel longer distances. In order to ensure passengers are satisfied, the extra distance travelled by them which is calculated through their detour ratio should be bounded by a threshold value. To illustrate this point consider the road instance shown in Figure 4a, where the vertices represent grid cells and the edge weights determine the distance between different grid cells. Assume the detour ratio is 1.5. The driver is at node $g_1$ and wants a recommendation mechanism from that point.

There are 3 paths that can be recommended between the grid cells $g_1$ and $g_6$: $P_1 = \{g_1, g_2, g_3, g_4, g_6\}$, $P_2 = \{g_1, g_5, g_6\}$ and $P_3 = \{g_1, g_7, g_8, g_6\}$. The efficiency of a path is determined by calculating the expected number of requests that can arrive in a path. This is done through the summation of edge weights $w^Q_{ij}$ from all the vertices $i$ in the path $P$ i.e, ($i \in P$) to the set of their forward nodes ( $j \in \mathcal{F}$) which lie on the path to the destination. Mathematically, it is represented as:

$$|P_{G^Q}| = \sum_{i \in P} \sum_{j \in \mathcal{F}} w^Q_{ij} \tag{6}$$

For example, in Figure 4b the expected number of requests in path $P_1 = \{g_1, g_2, g_3, g_4, g_6\}$ is $w^Q_{(1)(2)} + w^Q_{(1)(3)} + w^Q_{(1)(4)} + w^Q_{(1)(6)} + w^Q_{(2)(3)} + w^Q_{(2)(4)} + w^Q_{(2)(6)} + w^Q_{(3)(4)} + w^Q_{(3)(6)} + w^Q_{(4)(6)} = 0 + 0 + 0 + 2 + 1 + 1 + 0 + 1 + 0 + 1 = 6$. Thus we can say path $P_1$ expects 6 riders on its way. In a similar manner, path $P_2$ expects 3 riders on its way, and path $P_3$ expects 4 riders. The path $P_1$ contains the highest number of expected requests but it can't be recommended by the platform as it violates the detour ratio of some riders on board. This can be seen clearly followed through the detour ratio of rider 1 who has to travel from grid cell $g_1$ to $g_6$ and if it follows path $P_1$, the distance travelled by him will be 8, whereas the distance of the shortest path between grid cells $g_1$ and $g_6$ is 4 (see Figure 4a). The detour ratio of this rider is $\frac{8}{4}$ which is greater than the threshold value of 1.5. Thus this path cannot be recommended by the model. The next path with the highest requests is $P_3$, and it is recommended by the model as it does not violate the detour constraints of any of the riders and is efficient. Path $P_2$ also does not violate the detour constraints of any of the riders, however, it contains few requests than $P_3$, and is not recommended by the system. From this example, we can follow that although the maximum request path contains 6 requests, it cannot be followed by the driver as it violates the detour constraints of some riders in the vehicle.

Thus, the proposed model needs to select a path from the request graph that maximizes the expected number of requests and at the same time keeps a constraint on distances travelled by the riders. Mathematically, the problem is formulated as:

$$P^*_{G^Q} = \underset{P_{G^Q}}{\arg\max} \ \mathbb{E}[|P_{G^Q}|] \tag{7}$$

subject to

$$\forall_{o_s \in V} \alpha(P_{G^R}, o_s, o_d) \leq t_d \tag{8}$$

Eq. (7) represents the objective function for evaluating the efficiency of the proposed model, which is to choose a path $P$ from the request graph $G^Q$ that has the highest number of anticipated passengers. While maximizing the expected requests in the path there is a constraint at each step, specified by Eq. (8) which states that the detour ratio of all the rider requests should be bounded by a threshold value ($t_d$) in order to ensure the passengers are satisfied.

### 4.2 Problem hardness and approximations

In order to improve the efficiency of the system, the proposed model recommends routes with the highest passenger demand that satisfies the detour constraints of all the riders in the vehicle. However, recommending routes with the highest demand is NP-Hard as the Longest Path problem is reducible to it [11]. Due to the NP-Hardness of the problem, optimal algorithms cannot be applied to maximize the efficiency of the system. To solve the problem at hand effectively, there needs to be some pattern in the underlying structure of the problem. Previous studies have analyzed the trajectories of drivers and found that the 80% of trajectories have at least 80% of forward edges [28]. An edge $(u, v)$ in the graph is considered forward when the shortest path from node $v$ to the destination node $o_d$ is less than the shortest path from node $u$ to $o_d$. These findings suggest that drivers tend to move progressively towards their destination at each stage. Considering this fact, we re-design the request subgraph and select the set of forward edges only. This results in the creation of a Directed Acyclic Graph (DAG), as explained next.

### 4.3 Proposed Solution

In this subsection, we will describe the working of the proposed model.

*4.3.1 Creation of DAG.* The request graph $G^Q$ is a complete graph with requests between every pair of grid cells. We reduce it to a set of forward edges with a view to reduce its complexity and solve it in polynomial time. The point from which the driver wants the recommendation mechanism is considered the starting point or source node ($o_s$) of the graph and the DAG needs to be constructed from there onwards. We apply the greedy strategy and consider the node that has the maximum requests from the source node ($o_s$) as the destination node ($o_d$) of the graph. In this step, we have looked over the whole search space and selected the destination with the highest request value from the source node. Now, there is a ride with the source ($o_s$) and destination ($o_d$) fixed and we need to create DAG between these nodes. To do so, the proposed model selects a set of *forward* edges between *directly connected* grid cells. The forward edges are selected based on the results from the previous studies which show that the 80% of trajectories have more than 80% of forward edges [28]. However, all the forward edges of the request graph cannot be placed in DAG, rather the forward edges between the directly connected grid cells are selected. This is because the path is a sequence of connected nodes and if the proposed model will consider the whole search space from a node, it may find the highest requests to a point in the graph that is not directly connected to it. It will then again have to select the set of directly connected points between these nodes i.e, it will have to repeat the same procedure multiple times till it finds the directly connected nodes in a path. That is why we allow our model to see the destination of all requests from the source once (when selecting the destination node $o_d$) and keep the path connected after that.

In order to understand this point consider Figure 5 which shows the DAG of the request graph in Figure 4b. $g_1$ is the source point and the highest request point from it is $g_6$, which is considered as the destination point of the graph, and the DAG is constructed between these nodes. In the DAG, the edge between nodes $g_2$ and $g_4$ is not allowed, as these nodes are not directly connected (which can be seen through the road network instance shown in Figure 4a). If we would have allowed an edge between these nodes, then the path would not be feasible as the set of nodes in it were not adjacent, and we would have had to look for the directly connected nodes between them. To avoid this complexity, the proposed model looks over the whole search space from the source node when selecting the destination $o_d$ and thereafter selects the adjacent nodes whose source and destination are just one grid cell away.

Keeping these points in consideration, we redesign the request subgraph $G^Q$ to make it DAG $G^{Q'} = (V_{G^{Q'}}, E_{G^{Q'}})$ with the following constraints:

$$\begin{aligned} V_{G^{Q'}} &= \{v \in V_{G^Q} \mid |SP(v, o_d)| \leq |SP(o_s, o_d)|\} \\ E_{G^{Q'}} &= \{(u,v) \in E_{G^Q} \mid u, v \in V_{G^{Q'}},\ (u,v) \text{ is a forward edge} \\ &\qquad \text{and } u, v \text{ are connected}\} \end{aligned} \tag{9}$$

*4.3.2 Dynamic Programming framework.* In the previous subsection, we described a method to create a directed acyclic request graph. The acyclic nature of graphs provides us with a framework to solve the problem at hand efficiently. Although maximizing expected requests in a path is NP-Hard on graphs, it can be solved in polynomial time on directed acyclic graphs [19]. Given a request DAG, with two edges $(d_1, d)$ and $(d_2, d)$ to destination node $d$, the maximum request path (MRP) from source node $s$ to $d$ can be calculated as $max\ \{MRP(s, d_1) + w^{Q'}_{(d_1,d)}, MRP(s, d_2) + w^{Q'}_{(d_2,d)}\}$. Since this problem can be broken down into sub-problems, we apply dynamic programming (DP) to solve it efficiently. However, DP cannot be directly applied to find the optimal routes for drivers. This is because our problem involves two subgraphs and we have to recommend a path that maximizes the count of requests (determined by $MRP(s, d)$) through $G^{Q'}$ while keeping a constraint on the distances travelled by the riders through $G^R$. This problem demands maximizing one resource when the other resource is bounded. It displays that all the paths are not feasible, rather the paths whose edges satisfy the detour constraints of all the riders who are in the vehicle can be followed by the driver. Thus, we prune the search space like [28] and select the feasible paths, while maximizing the expected number of passenger requests.

If we analyze the creation of DAG clearly, we can follow that the origin and destination of all the requests are just one grid cell away except that of the rider(s) which has to travel from the source node ($o_s$) to the destination node ($o_d$). This is because, in the initial step, we have looked over the whole search space to see the maximum requests from the source node. After that, we kept the path connected and checked the adjacent nodes whose source and destination are just one grid cell away. The detour ratio of adjacent nodes will not be violated as the proposed model will move directly between them, but the detour ratio of rider(s) who has to travel from source to destination can be violated. Thus we need to take into account the detour ratio of this rider who has to travel from $o_s$ to $o_d$ when determining the feasibility of the paths.

The paths that satisfy the detour constraints of the rider $a$ who has to travel from $o_s$ to $o_d$ are considered feasible. Let $e_{ij} = (v_i, v_j)$ be an edge between the grid cells $v_i$ and $v_j$ in the path. The distance travelled by the rider $a$ till the vertex $v_i$ is denoted as $d_i$, and the distance between grid cells $v_i$ and $v_j$ is represented by $x$. The path $P$ is considered feasible if the detour ratio of the rider $a$ through all the edges in the path is less than threshold $t_d$ i.e,

$$\forall_{e_{ij} \in P_E} \frac{x + d_i + SP(v_j, o_d)}{SP(o_s, o_d)} \leq t_d \tag{10}$$

Here, $P_E$ is the set of edges between connected grid cells in path $P$. For instance, the edge set for path $P = \{g_1, g_5, g_6\}$ is $P_E = \{(g_1, g_5), (g_5, g_6)\}$.

To understand the feasibility of a path, consider Figure 5 in which we assume driver is at grid cell $g_1$ which is the source $o_s$ and it has 2 requests whose destination $o_d$ is $g_6$. There are 3 paths between nodes $g_1$ and $g_6$. $P_1 = \{g_1, g_2, g_3, g_4, g_6\}$, $P_2 = \{g_1, g_5, g_6\}$ and $P_3 = \{g_1, g_7, g_8, g_6\}$. Of these paths, $P_2$ is the shortest path between $g_1$ and $g_6$, and it is feasible as it will not violate the detour constraints of the rider $a$. We need to determine the feasibility of the other two paths $P_1$ and $P_3$. The feasibility of a path is checked by determining the feasibility of all the edges in the path. The first edge in the path $P_1$ is $e_{12} = (g_1, g_2)$. The distance $d_i$ travelled by the rider $a$ till $g_1$ is 0 as the rider wants a ride from that point and $x$ is the distance that rider $a$ can travel between grid cells $g_1$ and $g_2$. For the edge $e_{12}$ to be feasible, the distance $x$ should satisfy the detour constraint specified by Eq. (10) and $\frac{x+0+5}{4} \leq 1.5$. Here the length of the shortest path from $v_j$ which is $g_2$ to destination node $g_6$ is 5, and the length of shortest path between $o_s$ and $o_d$ is 4. The constraint in the above equation states that the edge $e_{12}$ is feasible if the distance between grid cells $g_1$ and $g_2$ is less than 2, but as can be seen, through the road graph displayed in Figure 4a, the distance between grid cells $g_1$ and $g_2$ is 3. Thus this path is not feasible as the detour ratio of rider $a$ is violated by travelling through its edge $(g_1, g_2)$. Similarly, we can check that the path $P_3$ is feasible as all the edges in it satisfy the detour constraint of rider $a$.

Now, that we have defined the feasible paths, we apply dynamic programming to the graph. Let $e_{v,k}$ denote the expected number of passengers that can be paired in a vehicle from node $v$ within the length $k$, without violating the detour constraints of any of the riders. Mathematically, it is defined as:

$$e_{v,k} = \max_{\forall P_{G^{Q'}} \in P_{G^{Q'}_{(v,k)}}} \{E[|P_{G^{Q'}}|]\} \tag{11}$$

where $P_{G^{Q'}_{(v,k)}} = \{P_{G^{Q'}} \mid |P_{G^{Q'}}| = k, P_{G^{Q'}}$ is a path from $o_s$ to $v$ in $G^{Q'}\}$. DP is initialized as:

$$e_{v,k} = \begin{cases} 0 & if \;\; v{=}o_s \; k{=}0 \\ -\infty & otherwise \end{cases} \tag{12}$$

The recurrence is defined as:

$$e_{v,k} = \max_{\forall (u,k') \in N_k(v)} \{e_{u,k'} + w^{Q'}_{(u,v)}\} \tag{13}$$

where $N_k(v) = \{(u, k') | (u, v) \in G^R, k = k' + w^R_{(u,v)} \leq \alpha_r\}$ is the set of the incoming node-length pairs that result in a path of length $k$ without violating the detour constraints $(\alpha_r)$ of the rider(s) who want to travel from $o_s$ to $o_d$. The recurrence is calculated through memoization where the incoming nodes and their path lengths are stored. These nodes are used for calculating the paths to the destination.

Table 6 displays the $e_{v,k}$ values of grid cells in DAG defined by Figure 5 when route needs to be recommended from $g_1$ to $g_6$. The $e_{v,k}$ values of grid cells $g_2, g_3$, and $g_4$ are not calculated as the detour ratio of rider who has to travel from $g_1$ to $g_6$ gets violated by travelling through the edges between those grid cells (as was explained in the previous example). All the other edges are feasible, and the $e_{v,k}$ values of their vertices are displayed in the table. Let us consider calculating the $e_{v,k}$ value of grid cell $g_6$. The incoming neighbor-length pair to the grid cell $g_6$, which includes the neighbors that don't violate the detour constraint of the rider who has to travel from $g_1$ to $g_6$, are defined by the ordered pair $(u, k)$ where $u$ is the incoming grid cell and $k$ is the distance travelled till node $u$. These include $N_k(g_6) = \{(g_8, 3), (g_5, 2)\}$. The $e_{v,k}$ value of $g_8$ is the expected number of passengers till node $g_8$ and it is 2 (as there is 1 request from $g_1$ to $g_7$ and 1 from $g_7$ to $g_8$). Similarly, $e_{v,k}$ value of $g_5$ is 0 as there is no request towards it. Since both the nodes are feasible, the proposed approach will select the node whose summation of $e_{v,k}$ value and the edge weight $w_{u,v}^{Q'}$ is maximum i.e, $max(e_{g_8} + w_{(g_8,g_6)}^{Q'}, e_{g_5} + w_{(g_5,g_6)}^{Q'})$. Putting in their values, we get $max(2 + 0, 0 + 1)$ which is 2 and it is reached through node $g_8$. The $e_{v,k}$ value, along with the node through which it is reached is stored in the rows 2 and 3 of Table 6. These rows are used for backtracking from the destination node to the source node in order to get the optimal path in DAG. In this example, we continue backtracking from node $g_8$ till the source node $g_1$ is reached, to get the optimal path as $P^* = \{g_1, g_7, g_8, g_6\}$.

*4.3.3 Ride-order acceptance in the recommended route.* The route recommendation model proposed above recommended the routes with the highest number of expected passenger requests. Once a route is recommended, the driver must then determine which passenger requests to accept. The driver can only accept requests that comply with the vehicle's capacity constraints and the detour constraints of passengers already in the vehicle. Let's assume the vehicle capacity is 3 and the detour threshold is 1.5. In order to understand how to take the ride orders without violating the capacity and detour constraints, let's see how the requests are taken after the optimal route $P^* = \{g_1, g_7, g_8, g_6\}$ was recommended to the driver in the previous subsection. We assume 6 requests arrive dynamically, of which 2 have to travel from $g_1$ to $g_6$, 1 has to travel from $g_1$ to $g_7$, 1 has to travel from $g_7$ to $g_1$, 1 has to travel from $g_7$ to $g_8$, and 2 have to travel from $g_8$ to $g_6$ (see Figure 7). When the driver is at its initial position $g_1$ capacity and detour constraints are not violated since there are no passengers on board yet and the vehicle is empty. So all 3 requests (2 from $g_1$ to $g_6$, and 1 from $g_1$ to $g_7$) are taken by the driver, and it follows the recommended route and moves to grid cell $g_7$ which has two requests of which one has to travel from $g_7$ to $g_1$, and the other has to travel from $g_7$ to $g_8$. If the driver takes the request from $g_7$ to $g_1$, two paths can be followed by him. The first path is $P_1 = \{g_1, g_7, g_1, g_7, g_8, g_6\}$ in which the driver first drops the passenger who is at $g_7$ to his destination $g_1$ and then travels towards $g_6$. If the driver follows this path, the detour ratio of passengers who have to travel from $g_1$ to $g_6$ will be the ratio of the length of the path followed by the vehicle between grid cells $g_1$ and $g_6$ (which is equal to the length of $P_1$ and is 9 as can be seen through Figure 4a) to the length of the shortest path between $g_1$ and $g_6$ (which is 4). This ratio is $\frac{9}{4}$, which is greater than the threshold value of 1.5. Therefore, this path cannot be followed by the driver. The other path is $P_2 = \{g_1, g_7, g_8, g_6, g_5, g_1\}$, in which the driver first drops the passenger who has to travel from $g_1$ to $g_6$ and thereafter drops the passenger who has to travel from $g_7$ to $g_1$. This path will not violate the detour ratio of passengers who have to travel from $g_1$ to $g_6$ as the shortest path is followed between their source and destination points. But it will violate the detour ratio of the passenger who has to travel from $g_7$ to $g_1$. Its detour ratio will be $\frac{9}{2}$ which is much greater than the threshold value of 1.5. So the request from $g_7$ to $g_1$ is not taken as both the paths $P_1$ and $P_2$ violate the detour constraints of passengers. The other request has to travel from $g_7$ to $g_8$ and it can be taken by the driver as its path is the subpath of the passengers who have to travel from $g_1$ to $g_6$ and both of the requests are following their shortest paths. Similarly, when the driver reaches grid cell $g_8$ there are 2 requests that have to travel to $g_6$, and the driver can only take 1 request as it already has 2 passengers in the vehicle (who have to travel from $g_1$ to $g_6$). This request from $g_8$ to $g_6$ does not violate the detour ratio of passengers already in the vehicle as its path is a subpath of the passengers who have to travel from $g_1$ to $g_6$, and it is taken by the driver. This is how orders are accepted within the recommended route.

| $v$ | $g_1$ | $g_5$ | $g_7$ | $g_8$ | $g_6$ |
|---|---|---|---|---|---|
| $e_{v,k}$ | 0 | 0 | 1 | 2 | 2 |
| $\rightarrow$ | – | – | $g_1$ | $g_7$ | $g_8$ |

Fig. 6. DP Table corresponding to DAG shown in Fig. 5

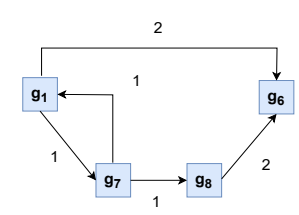

Fig. 7. Request order graph

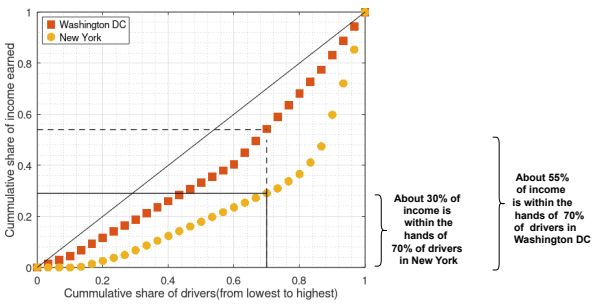

Fig. 8. Income distribution in New York and Washington DC

## 5 FAIR AND EFFICIENT ROUTE RECOMMENDATION SYSTEM

In the previous section, we described the working of a model that optimized system efficiency. While this model is used for increasing the customer base of platforms, it is unfair as it recommends routes based on the location of a driver. In this section, we will incorporate the fairness principles into the design of efficient route recommendation system described above.

### 5.1 Why Fairness?

Fairness is defined as the absence of any prejudice or bias towards an individual or a group based on their inherent or acquired characteristics. In simple words, we say an algorithm or data is unfair if it discriminates against individuals based on their features like gender, disability, or the place at which they are located. This issue has gained significant attention from the research community in recent years and it is mainly due to the prevalence of bias in algorithms ranging from Artificial Intelligence to recommendation systems. For example, in ride-hailing platforms, drivers working for similar hours may get different incomes based on their location (sparse or dense in requests), resulting in income disparities. If drivers continue to face discrimination, they may leave for other platforms, leading to a decrease in supply and a fare increase. This, in turn, may cause riders to switch to other platforms, creating a negative cycle that can harm the long-term sustainability of the system. Therefore, it is crucial to ensure fairness is satisfied over the long run to maintain the stability of the system.

### 5.2 Is the outcome of dynamic programming based model fair?

The dynamic programming framework is based on the utilitarian criterion of welfare economics which increases the efficiency of the system, by maximizing the cumulative utilities of the drivers. The driver is recommended a route that has the maximum number of riders and the additional distance travelled is bounded. However, if the driver is in a low-demand area, he may be left without a rider due to the lack of nearby passengers. Contrarily, the drivers in high-demand areas will receive the best recommendations and generate good income. This leads to a significant income disparity among the group of drivers, even if they are working for similar hours and providing the same level of service. In order to understand clearly how the algorithm discriminates against individual drivers, consider Figures 4a and 4b in which we assume there is a set of 2 drivers located at grid cells $g_1$ and $g_5$. The algorithm will recommend the routes $R_1 = \{g_1, g_7, g_8, g_6\}$ to driver 1, and route $R_2 = \{g_5, g_6\}$ to driver 2. By using Eqs. (2),(3) and (4), and the values of parameters in the equations as $b_f = \$2.55$, $m_f = \$7$, $f = \$1.75$, $f' = \$0.35$ and $c = \$0.13$, driver 1 will get a utility of \$27.5 and driver 2 will get \$6.74. From this example, we can follow that even though drivers 1 and 2 are just one grid cell away, they receive different recommendations and their generated income has a lot of differences. This bias is caused at the algorithmic level and it can affect the functioning of the system over the long run. To ensure that the system is fair and equitable, it is important to consider the needs and interests of drivers. We propose to remove this discrepancy by incorporating the fairness criterion in our route recommendation system.

### 5.3 Incorporating Fairness

Algorithmic fairness for ride-hailing platforms states that irrespective of other attributes, drivers working for similar time frames should get the same income. We need to consider the design of a route recommendation system that in addition to the constraints of ridesharing network, takes into account this notion of fairness and returns a set of routes for drivers that will provide them with similar income for similar time schedules. Before we go further into the design of this system, we need to ask the following questions.

**1) How do we incorporate fairness into our proposed model?**

In this subsection, we will examine the integration of a fairness metric into the objective function of the proposed system. In order to incorporate fairness, the algorithm needs to be designed in a way that provides comparable income to the drivers who work for similar time periods. The utility function defined in the previous section which is based on the utilitarian criterion, and reflects the cumulative income of drivers cannot be directly applied to optimize fairness. This is because different drivers may have worked for different time frames and if we equalize the cumulative utilities of drivers, the resultant distribution will provide all the drivers with the same income irrespective of their working hours which is clearly unfair. The metric that we consider using is *utility per hour* which normalizes the income of users for an hour and our proposed system will try to equalize this quantity for different drivers. Any time unit like hours, minutes, etc can be used for normalizing the income, but we propose to use hours as it provides a well-bounded framework that is neither too long nor too short.

After normalizing this metric, we can come across different incomes for different drivers. There may be some drivers with higher income and others with low or medium income. To remove this discrepancy, the proposed model needs to provide a higher valuation in assigning routes to the drivers that have low income. In order to do that, we need to modify our objective function and incorporate a function that provides higher value to low-income people, i.e, this function follows the *law of diminishing returns* according to which a value added to a larger set does not yield much of value in comparison to a smaller set. One of the well-known sets of functions that are known to follow the law of diminishing returns is the concave function.

Any concave function can be used with our objective function, but we propose to use the *log* function as it converges to the well-known notion of fairness called the *proportional fairness* which has been used to model fairness issues in recent literature. Thus the objective function is re-framed to model fairness and it is mathematically represented as: $max\ \sum_{i=1}^{n_d} \log(u_i(r, d, c, n))$.

After incorporating the fairness metric into the proposed system, we need to determine the interval over which fairness can be achieved, which leads to the second question.

**2) Can fairness be satisfied over a single run? If not, how much do we deviate from it over a single iteration?**

In this subsection, we will analyze whether it is possible to achieve fairness in a single instant, or if it needs multiple iterations to settle down. Moreover, if the model requires multiple iterations, how much deviation can be expected in a single iteration?

We can follow through the intuitive reasoning that no matter which permutation of routes is assigned to drivers, these routes will always be beneficial for some at the cost of others. This provides us with an indication that the model cannot ensure fairness in a single run, but it needs to be *amortized* over multiple rounds. Thus if any driver is assigned with the best route at time instant $t$, at the next time instant $t + 1$ another driver should be given priority over the allocation of the best route. In this manner, the outcome of the system can turn out to be fair over time.

When fairness is amortized, the deviation of its spread from the ideal fair case should be quantified i.e, at each instant of time we should know how much our proposed model diverges from the ideal fair distribution. It will provide us with a bound over the worst performance of the model. In order to quantify this measure, we propose to use the Gini index and see how it varies over time. It is a measure of inequality and its value lies between 0 and 1, with 0 denoting the case of perfect equality and 1 determining perfect inequality. We determine its value for each round and see how it varies over time, in order to determine the performance of the proposed system.

After determining the interval over which fairness can be achieved, we need to determine if fairness is correlated with the demand at that point in time, which leads towards the third question.

**3) Can we reduce algorithmic bias for all hours irrespective of the demand at that point in time?**

In this subsection, we will address the third question pertaining to the relationship between demand and fairness. Specifically, we aim to investigate whether it is possible to reduce bias without taking into account the demand at the point in time when drivers are working. It has been established that there exists a strong correlation between passenger demand at a given time period and the income of drivers [20]. If the demand is low during a particular time period, regardless of the fairness algorithm employed, it is unlikely to equalize the earnings of a driver working during this period with that of a driver working during high-demand hours. Therefore, when designing fair algorithms, it is crucial to incorporate the demand factor and consider it as a weight over the time period when routes are recommended. The objective function gets modified accordingly to incorporate this weight and is re-framed as:

$$max \sum_{i=1}^{n_d} w_i \cdot \log(u_i(r, d, c, n)) \tag{14}$$

This function is a variant of proportional fairness, called *weighted proportional fairness*, where the weight $w_i$ is added to each driver that corresponds to the passenger demand in its working hours. The demand can be estimated by the average income of drivers working in that hour. If the demand for a time period is high, the drivers working in that time frame will have high income and vice versa. Thus the proposed problem after incorporating fairness is re-framed and is defined by Eqs. (8) and (14). The problem demands maximizing the expected requests of the system while providing similar income to the drivers working for the same time periods, keeping in consideration the detour constraints of the riders. As the objective function is concave with linear constraints, its optimal solution can be found through standard Lagrangian multipliers and is equal to $u_i^* = \frac{w_i}{\sum_{j=1}^{n} w_j}$. The use of this notion of fairness ensures equal utility per hour for all the drivers weighted by the demand over the point in time when the drivers are working. To incorporate this notion in our proposed system, and provide a higher valuation to low-income drivers, we need to remove the bias that exists due to the locational sparsity of some drivers and provide them with routes that will have high request intake in the present as well as in the future. Moreover, there is another element in the recommendation system that occurs due to simultaneous requests from a set of competing drivers. When a set of drivers are recommended a route simultaneously, the driver who gets it at the earliest will receive the best, which can further enhance the existing disparities in the system. Thus, we can say there are two aspects of fair recommendation: the temporal aspect which considers the simultaneous recommendation from a set of drivers, and the spatial aspect which takes into account the current location of the driver. The proposed approach takes both of these aspects into consideration and provides a recommendation system that removes the differences in space and time. To remove the differences in time, we develop a priority-based algorithm that provides higher priority in assigning routes to low-income drivers and removes the differences in simultaneous recommendations. For location-based differences, which are the main reason for the income disparity, the system relocates the drivers towards areas that have high request intake. We will describe their working in detail in the next subsections.

*5.3.1 Priority-scheduling for simultaneous recommendations.* The priority-based algorithm is proposed to decrease the income differences among a set of drivers who are seeking recommendations from the platform simultaneously. The algorithm takes as input the utilities of drivers and assigns a top priority to the drivers with low utilities. We assume top priority provides a higher valuation to drivers and gives them earlier chance to get a route. After the drivers are assigned priorities, the algorithm recommends routes to them in accordance with their increasing priorities. In this way, drivers with low income are recommended routes at the earliest which provides them with a higher chance of getting riders on their way. This notion of priority takes into account the temporal dimension of recommendation and recommends a set of routes to competing drivers in a simultaneous manner keeping in consideration the fairness of the system. It is based on the fact the fraction of routes recommended to different drivers may be similar and the driver who gets the recommendation at the earliest will take all the riders on that route and the others will be left with a few riders. This concept holds particular significance when there is a surge in passenger demand, resulting in an increased supply of drivers, and a relatively uniform allocation of routes among them. The dynamic programming-based algorithm recommended the routes in a First Come First Serve manner, which can enhance the existing inequalities in the system. Our proposed model will remove the unfairness in time by assigning routes on priority to low-earning drivers, which provides them with preference in the assignment of passengers on common routes and marginalizes their utilities per hour within the bounds of high-earning drivers. This results in the uniform distribution of income in the simultaneous recommendations from a set of drivers.

*5.3.2 Relocation for location-based unfairness.* The second aspect of the fair route recommendation system is the mitigation of disparity due to the spatial constraints of drivers, where some drivers are placed in fine positions which provides them with better opportunities to get passengers, in comparison to others. In order to overcome this issue, we propose to relocate the drivers with low income to places where they will have a high probability of getting riders. However, it is not feasible to relocate all the drivers since the platform comprises a subset of top-earning drivers who possess knowledge of the routes that offer them a higher probability of obtaining more riders. This finding is supported by both prior research [30] and our experimental analysis, which indicates that income in ride-hailing platforms is predominantly concentrated among the top 30% of drivers.

Figure 8 shows the Lorenz curve of the income distribution in the ridesharing platforms. As can be seen clearly, the top 30% of people have 70% of income in their hands, in New York City and 45% in Washington DC. From this data, we can conclude the top 30% of drivers do not need to be relocated as they earn their share of income in a fine manner, rather the bottom 70% of drivers ask for relocation to keep up with the system performance and enhance its stability. Keeping this fact in place, we relocate the bottom 70% of drivers towards areas that have high passenger demand. The relocation will provide them with better opportunities to fetch passengers and will result in uniform distribution of income among the different set of drivers.

In order to design the relocation policy, the proposed model categorizes the drivers based on their utilities per hour and splits them into two groups: the bottom 70% and the top 30%. As the top-performing drivers do not need to be relocated, they are provided with recommendations directly through the dynamic programming framework. However, the low-earning drivers are relocated to the areas with a high request intake in order to balance the income differences among different groups of drivers. This is done through the creation of a window around the driver's current location. The driver's starting location is taken as the middle point of the window and its adjacent 8 grid cells are taken to determine the best location where it can be positioned before getting the route recommendation through the dynamic programming approach. While applying relocation, a destination-aware policy is framed and the areas are allocated to the driver considering the destinations of requests in that area. This future-aware policy ensures drivers are relocated to the areas that will have a high request intake at present as well as in the future. To apply this mechanism, the proposed model sorts the locations within the window by their request count and checks the destination of the requests that want to travel from that location. If the requests from the highest request area in the window (first in the sorted list) land in areas that will have at least one request in future, the driver is relocated to that area. However, if the destination of requests from that location leads the driver to a sparse area that is not expected to have any requests in the future, the driver is not relocated there. Rather the next location in the sorted list with the higher passenger intake is checked for its sparsity, and this process continues till the requests from the location land in dense areas. The destination-aware policy ensures that the driver is relocated to a place that will have a high request intake in present as well as in the future. It provides them with a route that brings them through the shortest path to the area where requests will come in near future and prevents them from staggering in areas with low requests. In this way, relocation provides drivers with better opportunities to get riders and mitigates the disparity due to location differences between drivers.

Thus, we can say the implementation of a fair route recommendation system that integrates priority and relocation-based frameworks effectively evens out income distribution among drivers operating within similar time frames. The efficiency maximization framework considered the current location of the driver, and not their accumulated income for recommending routes to drivers which resulted in varying income among them even if their working hours aligned. The fairness design proposed considered the driver's accumulated income and relocated them from sparse areas to the areas that are dense in requests. This approach increased the probability of drivers getting matched with a ride request which apart from improving fairness also improved the utility of the platform. The key reason behind this enhancement lies in the strategic placement of drivers in high-supply areas, which reduces the difference between demand and supply and results in an overall improvement in the platform's utility.

Algorithm 1 presents the working of a fair and efficient route recommendation system. It sorts the driver according to their utilities per hour (line 2) and assigns routes in accordance with their sorted utilities (line 3). For each driver taken in order of its priority, its source location is fetched (line 4). If the location is sparse, driver is relocated (lines $5-7$). The point with maximum requests from the source is selected as destination point $o_d$ and DAG $G^{Q'}$ is constructed between the source and destination points (lines $8-9$). The vertices in DAG are sorted by topological order which ensures the incoming neighbors of a node are processed before it (line 10). $e_{v,k}$ is calculated for each node (lines $14-17$) and the path is backtracked from destination node $o_d$ by following the maximum $e_{v,k}$ of nodes (lines $18-20$).

---

**Algorithm 1:** Fair and efficient route recommendation system

---

**Input:** Number of drivers $n_d$, Starting location of drivers $\{o_{s_i} \mid i \in [1, n_d]\}$, utility per hour of drivers $\{u_i \mid i \in [1, n_d]\}$, detour threshold $\alpha$ , request graph $G^Q$, road graph $G^R$

**Output:** Fair and efficient path $p^*$

1: $p^* = []$
2: $d_s \leftarrow$ Sort the drivers according to their utilities per hour
3: **for** *driver* in $d_s$ **do**
4: $\quad o_s \leftarrow$ Starting location of *driver*

5: **if** $o_s$ is sparse location **then**
6: $o_s \leftarrow$ Relocate *driver* to new location
7: **end if**
8: $o_d \leftarrow$ Node with maximum requests from the source node $o_s$
9: $G^{Q'} \leftarrow$ Construct DAG between $o_s$ and $o_d$
10: $\forall\, v \in V_{G^{Q'}}$, compute topological-order($v$)
11: $e_{o_s,0} \leftarrow 0$
12: **for** $j \leftarrow 1$ to topological-order ($o_d$) **do**
13: $V_j \leftarrow \{v \in V_{G^{Q'}} \mid topological - order(v) = j\}$
14: **for** each $v \in V_j$ **do**
15: For all neighborhood grid cells that do not violate detour constraints of rider(s) on board compute $e_{v,k}$
16: **end for**
17: **end for**
18: $e^* \leftarrow \max\limits_{\forall\, k}\{e_{o^d,k}\}$
19: $p^* \leftarrow$ Backtrack from $o_d$ with $e^*$ to get optimal path in DAG
20: return $p^*$
21: **end for**

### 5.4 Complexity analysis

The time complexity of the proposed approach depends upon the sorting of drivers, and the recommendation of routes to drivers through dynamic programming. Sorting the drivers is a log-linear process, taking $O(n_d \log n_d)$ time. After sorting, the proposed approach concurrently handles each driver, initiating the relocation of drivers with low utility per hour. The relocation operation has a time complexity of $O(l)$, where $l \subseteq |V|$ is the number of grid cells within which drivers are relocated. After relocation, DAG of a grid is created in $O(|V|)$ time. Thereafter, routes are recommended through dynamic programming, where the grid cells are initially sorted using topological sort in $O(|V|)$ time. Once the topological order is determined, each vertex is systematically processed to determine the maximum requests among the adjacent backward edges. The proposed model's grid representation ensures a maximum of 5 backward edges, making this step to be executed in $O(|V|)$ time. This restriction arises from the fact that, in the road network, each grid cell $v$ is connected to 8 neighboring cells. Out of these neighbors, a maximum of 5 nodes can lie on the backward path, while the remaining nodes are forward neighbors whose edge weights depend upon vertex $v$. To understand this, consider Figure 1 where $e_{v,k}$ of $g_9$ has to be calculated when route needs to be recommended from $g_8$ to $g_7$. In this case, the adjacent backward edge set of $g_9$ is $\{(g_1, g_9), (g_2, g_9), (g_8, g_9), (g_{15}, g_9), (g_{16}, g_9)\}$, which is used for updating its $e_{v,k}$ value. Thus the overall time complexity of the proposed approach is $O(n_d \log n_d + |V|)$.

## 6 EXPERIMENTS AND RESULTS

In this section, we determine through experimental evaluation that our proposed model is fair, efficient, and scalable, which makes it practical and sustainable for route recommendation systems in ridesharing platforms.

### 6.1 Experimental setup

The experiments are implemented in Python and carried out on a machine with Intel (R) Core (TM) i9-12900 CPU 2400 MHz and 32 GB RAM.

*6.1.1 Datasets.* The performance of our proposed model is evaluated on two real-world datasets generated by Washington DC and New York. The datasets were collected for the month of February 2017 and 2018 respectively. Each dataset is divided into grid cells of 1.24 *miles* with a time gap of 15 minutes. The rows of the dataset are of the form pick-up time, pick-up latitude and longitude, drop-off latitude and longitude, and passenger count. This data about passengers' origin and destination is fed as input to the GNN-based model described in [10], which predicts the number of requests that can arrive between any two locations within the next 15 minutes. The predicted data is fed as input to the request graph $G^Q$ which is used for recommending routes to the drivers.

The GNN-based model uses the data from previous hours to determine the contextual information of the place. If the previous hours have few requests it implies either the area is sparse in requests or there is congestion in the subsequent regions which has reduced the flow of requests to that area. Thus we can say the expected number of passenger requests in an area determines the contextual information of the place such as its congestion level which helps the route recommendation system to take these factors into account apart from the passenger demand at the place.

*6.1.2 Evaluation framework.* To evaluate the performance of our proposed model, we divide the dataset into training data and test data. The training dataset contains 75% of the data and the rest is used for testing purposes. The model is trained for 200 epochs on the Washington DC dataset and 100 epochs on the New York dataset, and thereafter its performance is evaluated on the test set. The data generated by the test set which includes the expected number of passengers between different grid cells is used as input by the route recommendation algorithm to provide optimal routes to the driver. After using the predicted data, the performance of the route recommendation system is measured by using the actual data of passengers' origin and destination of the corresponding place.

While evaluating the performance of the proposed model, the values of variables are set as: base fare $b_f = 2.55$, minimum fare $m_f = 7$, fare per unit of distance (in miles) $f = 1.75$ , fare per unit of time (in minutes) $f' = 0.35$, and cost (per mile) $c = 0.13$.

*6.1.3 Baselines.* The following baselines have worked on optimizing efficiency and/or fairness in ride-hailing platforms.

**SHARE [28].** It optimizes the efficiency of the system by recommending routes with higher demand.

**URP [25].** It is a route planning system that inserts a new request in the original route of the driver dynamically through its insertion operation.

**TESLA [17].** It optimizes the earnings of drivers and reduces the differences in their earnings through the design of matching algorithms and route recommendation system.

**PA [14].** It maximizes the efficiency and fairness of the system by providing priority recommendations to low earning drivers.

**Greedy algorithm.** It uses a greedy approach and selects the node with maximum requests at each point of time.

*6.1.4 Metrics.* We evaluate the performance of our proposed model through the following well-known metrics:

**Efficiency.** It is the cumulative sum of utility per hour of all the drivers, and should be maximized for optimal functioning of the system.

**Lorenz Curve.** It illustrates income inequality by plotting the cumulative percentage of total income earned by a given percentage of the population. The greater the deviation from a straight line, the higher the inequality among individuals.

**Gini coefficient.** It is a single number that determines the spread of the Lorenz curve from the line of perfect equality. The lower the value, the better the performance of the proposed model.

**Waiting time.** It is the time between passenger request and vehicle arrival.

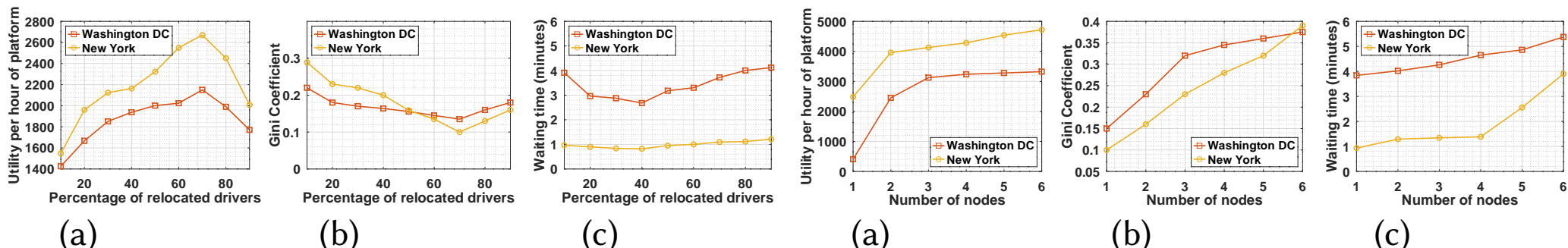

Fig. 9. Effect of driver relocation on (a) utility, (b) gini coefficient, (c) waiting time

Fig. 10. Effect of node count relocated on (a) utility, (b) gini coefficient, (c) waiting time

### 6.2 Results and Discussions

In this subsection, we will analyze the performance of our proposed framework on the metrics specified above. Firstly, we will see the effect of parameters on the proposed system, and thereafter, we will compare our model with the existing baselines.

*6.2.1 Parameters.* There are three parameters that affect the working of the proposed system. These include the detour ratio, the percentage of drivers that are repositioned to the new location, and the area within which they are repositioned. The detour ratio determines the extra distance that can be travelled by the passengers in the vehicle beyond their shortest path. Since the extra distance can be different depending on the preferences of passengers, we have set this value as a random number between 1 and 2.

**Percentage of relocated drivers.** The second parameter is the percentage of drivers that are relocated. Figure 9 shows the effect of the relocation of a fraction of drivers on the system metrics. With the increase in the percentage of drivers that are relocated, the utility of drivers is found to increase in New York as well as in Washington DC (see Figure 9a) till the 70% of drivers are relocated and decreases afterward. This increase in the utility is due to the increase in the service coverage of the ridesharing platforms through the placement of drivers in the high request-intake areas. The decrease in utility after the 70% of drivers are relocated is attributed to the fact that the top 30% of drivers contain a good share of income which is near about 70% in New York and 45% in Washington DC (see Figure 8). When these drivers are relocated to better areas it increases their income, but it results in a considerable drop in the earnings of the bottom 70% of drivers and decreases the efficiency of the platform.

Apart from the utility, the increase in the percentage of relocated drivers also has a positive effect on the fairness of the system till 70% of drivers are relocated, as can be seen through their Gini coefficient values displayed in Figure 9b. With the increase in the percentage of relocated drivers, the lower earning drivers are provided with locations with higher passenger demand which improves their income and decreases the Gini coefficient values. However, after 70% of drivers are relocated the further increase in percentage of relocation increases the Gini coefficient values. This is because the top 30% of drivers already have a higher share of income among them and if they are provided with better locations their income will improve which will enhance the already existing income disparities among drivers and lead to higher Gini coefficient values. Moreover, the Gini coefficient value is higher in New York than in Washington DC. The intuitive reasoning behind this analysis is the higher passenger demand in New York than in Washington DC which leads to higher utility in New York and also results in higher income differences among drivers which shows up in their Gini coefficient values.

Though fairness and efficiency are important parameters of the system, they cannot be increased at the cost of the waiting time of passengers. Figure 9c shows the effect of increasing the percentage of relocated drivers on the waiting time of the system. The increase in the percentage of relocated drivers leads to an increase in the waiting time of riders after the 40% of drivers are relocated, but this increase is linear and does not affect the performance of the system. This linear increase allows us to relocate the 70% of drivers without affecting the waiting time of riders considerably. The waiting time does not decrease initially till 40% of drivers are relocated since the drivers are shifted to areas with more demand which decreases the waiting time of passengers over those areas. But with the consistent increase in the percentage of relocation, there is an increase in the waiting time of the riders from low-demand areas which increases the average waiting time of the system. Further, the increase is found to be less in New York City than in Washington DC. This is primarily due to the higher demand in New York which makes drivers find the passengers within the relocated areas quickly. From this analysis, we conclude that relocating the bottom 70% of drivers results in a fair and efficient system without affecting the waiting time of riders considerably.

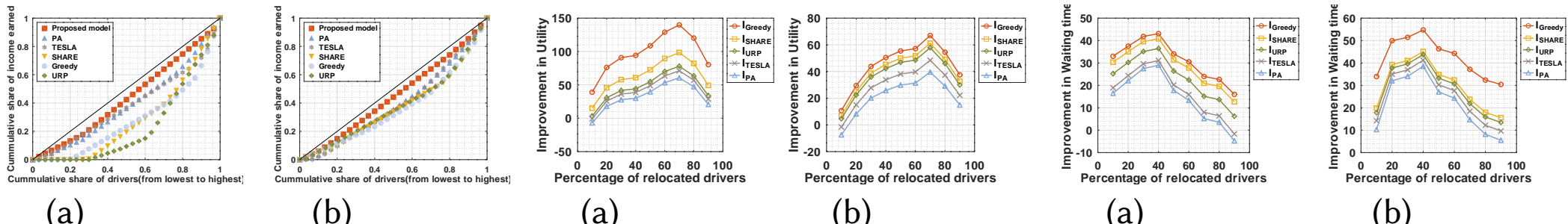

Fig. 11. Lorenz Curve comparison on (a) New York, (b) Washington DC dataset

Fig. 12. Utility comparison with relocated drivers on (a) New York, (b) Washington DC dataset

Fig. 13. Waiting time comparison with relocated drivers on (a) New York (b) Washington DC dataset

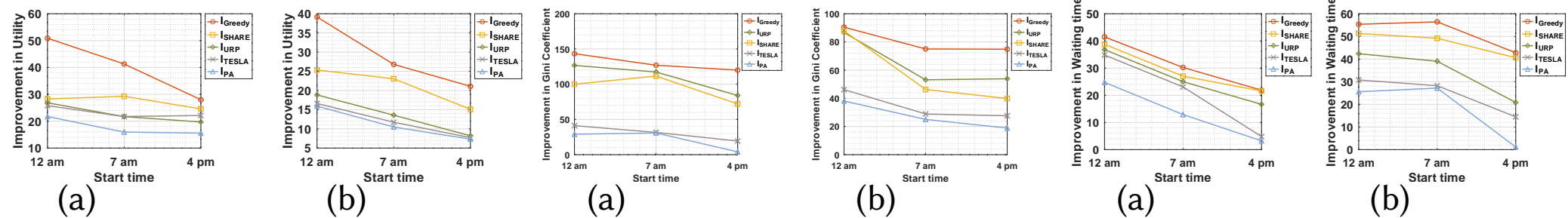

Fig. 14. Utility comparison across time frames on (a) New York, (b) Washington DC dataset

Fig. 15. Gini coefficient comparison across time frames on (a) New York, (b) Washington DC dataset

Fig. 16. Waiting time comparison across time frames on (a) New York, (b) Washington DC dataset

**Area of relocation.** The third parameter that determines the working of the proposed system is the area within which the drivers are relocated. Each grid cell is of size $1.24 * 1.24$ $miles^2$. If the drivers are relocated within the neighborhood 8 grid cells, which corresponds to the 1.24 *miles* area from each side, the system performance on the fairness is optimal. This observation is evident from Figure 10b, where the X-axis represents the number of nodes (grid cells) within which the driver is relocated and the Y-axis determines the Gini coefficient. This decrease in fairness can be attributed to the higher utility of the drivers on the increase of relocation area (see Figure 10a), which leads to higher income differences among them. Though the larger area offers better relocation options and increases the utility of the platform, it is not feasible due to an exponential increase in passenger waiting time (see Figure 10c), which leads to customer dissatisfaction with the system. Thus we have kept the default value of the area of location as neighborhood 8 grid cells.

*6.2.2 Comparison with the existing baselines and evaluation of its performance.* The performance of our proposed model is evaluated by comparing the percentage improvement ($I$) over each specified metric for all the baselines. Mathematically, improvement is defined as: $I = \frac{P-B}{B} \times 100$, where $B$ denotes the performance of the baseline method and $P$ denotes the performance of our proposed method.

Figures 11a and 11b show the performance of the proposed system and the existing baselines on the fairness metric: Lorenz curve. Lorenz curve displays the percentage of income that is concentrated within the hands of the population. As can be seen through Figures 11a and 11b, with the baseline approaches a higher fraction of income is concentrated within the hands of the top 30% of drivers in New York, and Washington DC. Our proposed approach overcomes this problem by relocating the bottom 70% of drivers to the areas with higher passenger demand which reduces the income disparity among the set of drivers working for similar time periods.

Even after incorporating the fairness metric the proposed model is able to optimize the utility of the platform, as can be seen through Figures 12a and 12b, where $I_{Greedy}$ , $I_{SHARE}$, $I_{URP}$, $I_{TESLA}$, and $I_{PA}$ denote the improvement of our proposed model over the baselines Greedy, SHARE, URP, TESLA, and PA. The intuitive reasoning behind this improvement is the increased service coverage of the platform through optimal relocation of a set of low-earning drivers who otherwise are found to cruise around without having a rider in their vehicle. By relocating them to better locations, the proposed system provides them with higher income which results in a decrease in discrimination and also increases the cumulative utility of the platform. It can be seen through the figure that the proposed model improves in performance even when a fraction of drivers are relocated. This is primarily because the proposed model utilizes the origin and destination of requests which results in efficient pairing of passengers in vehicle and increases their utilities. The existing studies [25, 28] have primarily applied demand prediction wherein these models have predicted the passenger demand in different areas and recommended optimal routes. However, the route recommendation for ridesharing platforms is efficient if the origin as well as destination of requests can be predicted beforehand as it will result in the efficient pairing of different passengers [10].

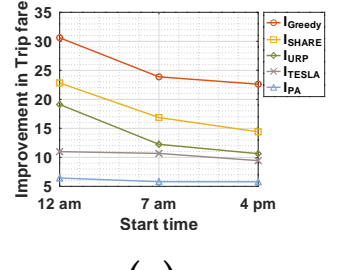

(a)

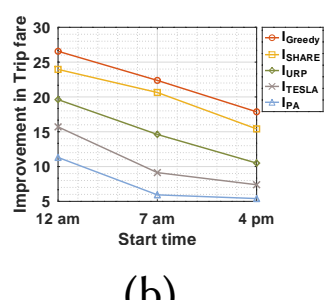

(b)

Fig. 17. Trip fare at different time frames in (a) New York, and (b) Washington DC

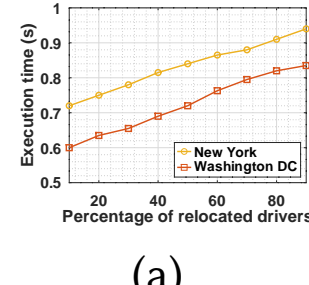

(a)

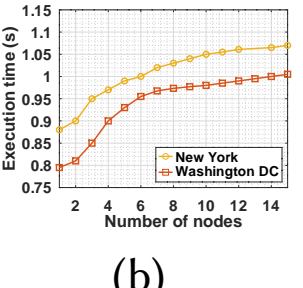

(b)

Fig. 18. Execution time with (a) relocated driver percentage, (b) nodes

The next important metric that should be taken into consideration is the waiting time of the passengers. Figures 13a and 13b show the improvement in the waiting time of the proposed system over the existing baselines. The proposed model reduces the waiting time in comparison to the existing baselines. With the existing baselines, drivers are provided routes that will have high passenger demand. However, if drivers are in a low-demand area, they will be recommended to the neighboring node with higher demand. Since the neighborhood node is also likely to have low demand, this process will continue until the driver reaches an area with demand. Conversely, in high-demand areas, there may not be enough drivers to serve all passengers, resulting in increased waiting times. Our proposed approach is able to overcome this issue by relocating the drivers in the low-demand areas through the shortest path to the areas that will have high customer intake which leads to a decrease in the waiting time of the riders, and an improvement in utility, and fairness of the system.

**Performance of the proposed model over different time frames.** Figure 14, 15 and 16 present a comprehensive overview of the performance of our proposed model during three distinct time intervals: 12:00 AM, representing nighttime when demand is relatively low; 7:00 AM, corresponding to the morning rush hours as people commute from home to the office; and 4:00 PM, signifying the evening rush hours when individuals return from their workplaces. These figures clearly demonstrate that the proposed model consistently delivers strong performance across all time frames. During the nighttime, when demand is relatively low, the performance of our proposed model is much higher than the existing baselines. This is primarily because the proposed model uses relocation and priority approaches to distribute the drivers across different areas and improves performance significantly. The performance drop observed during the morning and evening rush hours can be attributed to the heightened demand during these periods. During these peak hours, even the existing baselines perform well, resulting in more modest performance improvements. However, these studies consistently underscore the robust performance of our proposed model when compared to the existing baselines across various timeframes characterized by fluctuating demand levels. In addition to the previously discussed metrics, we've also measured customer satisfaction by evaluating trip fares across various time frames. Figure 17 visually presents the comparative enhancement in trip fares achieved by our proposed model when contrasted with the existing baseline models. Like other metrics, the increase in trip fares is most pronounced during nighttime, with a gradual decrease observed during the morning and evening rush hours. This reduction in fare improvement is primarily due to the increased demand during peak hours,

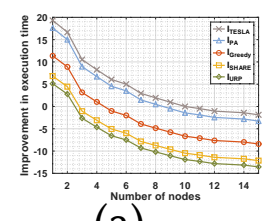

(a)

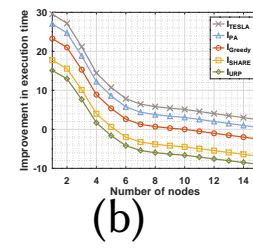

(b)

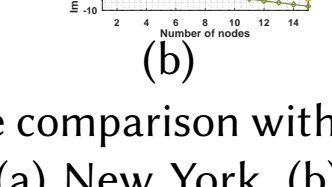

Fig. 19. Execution time comparison with increase in nodes on (a) New York, (b) Washington DC dataset

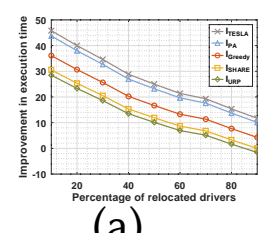

(a)

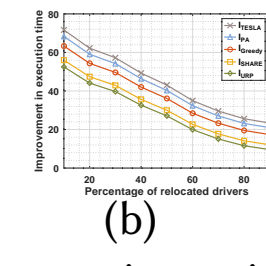

(b)

Fig. 20. Execution time comparison with increase in relocated drivers on (a) New York, (b) Washington DC dataset

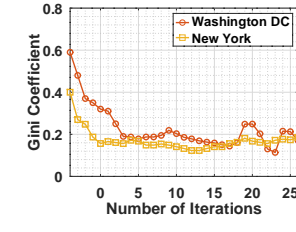

Fig. 21. Gini coefficient analysis

where the existing baseline models perform efficiently, resulting in more moderate performance enhancements compared to nighttime.

6.2.3 *Execution time.* The proposed model's execution time is assessed based on two key parameters: the number of nodes and the percentage of relocated drivers. Figure 18a depicts the execution time of the proposed model as the percentage of relocated drivers increases. It is evident from this figure that the execution time shows a linear increase with the percentage of relocated drivers. Notably, even when all drivers are relocated, the proposed model exhibits robust scalability, completing execution within 1 second. This underscores its effectiveness in adapting to varying percentages of relocated drivers.

Figure 18b illustrates the execution time of the proposed model with the increase in the number of nodes. Initially, the execution time experiences a notable rise alongside the growing number of nodes, followed by a gradual ascent. This trend can be attributed to the fact that a significant portion of drivers successfully locate passengers within a confined relocation area, leading to a marginal increase in execution time with the increase in number of nodes. This figure demonstrates that the increase in the relocation area does not increase the execution time considerably, indicating the proposed model's ability to scale to large road networks represented in the form of billion-sized graphs.

**Comparison of execution time with existing baselines.** We conducted a comparative analysis of the execution time of our proposed model against baseline approaches. Figures 19a and 19b illustrate the performance improvement of our proposed model as the number of nodes increases, using datasets from New York and Washington DC, respectively. With the increase in the number of nodes, there is a rise in the execution time of the proposed model due to the need to process a larger amount of data. The increase in execution time is more prominent than the efficiency maximization baselines - SHARE and URP, whereas it is smaller than the fairness baselines - TESLA and PA. This is primarily because the efficiency maximization baselines recommend routes with the highest passengers without considering the fairness frameworks like relocation and priority, which leads to a lower count of operations performed by them and shows up in lower execution time. The execution time is also higher than the greedy baseline since greedy looks for locally optimum choices and runs faster. Despite the slight increase in execution time as compared to efficiency maximization and greedy baselines, the proposed model still responds to any query within 1.1 seconds (see Figure 18b), ensuring real-time responsiveness in practical settings.

Apart from the increase in the number of nodes, we also compared the execution time of the proposed model with the increase in the percentage of relocated drivers as can be seen through Figures 20a and 20b. It's intuitive that as the percentage of relocated drivers increases, the improvement in execution time diminishes, given that a higher number of relocated drivers translates to longer processing times. However, our proposed model consistently outperforms existing baselines, even under scenarios where the majority of drivers are relocated. This is because our proposed model relocates the drivers within the neighborhood nodes while optimizing the platform utility, fairness, and waiting times simultaneously, which shows up in lower execution time and higher improvement. It displays that the proposed model is scalable and can be applied in real-time settings.

6.2.4 *Gini coefficient analysis.* Figure 21 shows the Gini coefficient values over different time frames. The model starts recommendation at 9:00 am which corresponds to the first iteration. Each subsequent reading is taken after 15 minutes and the last reading corresponds to the evening hour of 5:00 pm. This figure highlights three important conclusions. Firstly, we can observe that the performance of the proposed system improves over subsequent iterations. The Gini coefficient is initially high as the model starts recommendation, but it decreases considerably over the next iterations, indicating that the model's performance is improving over time. Second, a direct correlation between demand and fairness is observed; high Gini coefficients during morning and evening rush hours are attributed to supply-demand imbalances, while low-demand periods exhibit more equitable distributions resulting in a lower value of Gini coefficient. Lastly, this figure demonstrates that the proposed model does not achieve the ideal Gini coefficient of 0. This observation is further substantiated by examining Figures 11a and 11b, which illustrate that the performance of our proposed model does not reach an optimal level as it falls short of achieving the line of perfect equality. This is because no algorithm can completely account for the significant variability in demand and the diverse schedules of drivers, which ultimately leads to disparities [20]. In other words, drivers operate at different times, and demand varies considerably, making it challenging to achieve perfect equality.

## 7 CONCLUSION

This paper establishes that a fair and efficient route recommendation system can be developed for ridesharing platforms. The proposed approach does this by designing a priority and relocation algorithm that derives its principles from weighted proportional fairness. Further, it overcomes the scalability challenges posed by the NP-Hard nature of route recommendation, by reducing the search space to a directed acyclic graph, which is then solved using dynamic programming. Through this approach, the proposed model is able to jointly optimize the fairness and utility of the system without increasing the waiting time of the riders considerably. Moreover, the proposed approach is scalable to billion-sized graphs as it responds to any query within 1.1 seconds.

## REFERENCES

[1] Lu Chen, Qilu Zhong, Xiaokui Xiao, Yunjun Gao, Pengfei Jin, and Christian S Jensen. 2018. Price-and-time-aware dynamic ridesharing. In *2018 IEEE 34th international conference on data engineering (ICDE)*. IEEE, 1061–1072.

[2] Zhao Chen, Peng Cheng, Lei Chen, Xuemin Lin, and Cyrus Shahabi. 2020. Fair Task Assignment in Spatial Crowdsourcing. *Proc. VLDB Endow.* 13, 12 (2020), 2479–2492.

[3] Peng Cheng, Hao Xin, and Lei Chen. 2017. Utility-aware ridesharing on road networks. In *Proceedings of the 2017 ACM International Conference on Management of Data*. 1197–1210.

[4] Guang Dai, Jianbin Huang, Stephen Manko Wambura, and Heli Sun. 2017. A Balanced Assignment Mechanism for Online Taxi Recommendation. In *2017 18th IEEE International Conference on Mobile Data Management (MDM)*. 102–111.

[5] Nandani Garg and Sayan Ranu. 2018. Route Recommendations for Idle Taxi Drivers: Find Me the Shortest Route to a Customer!. In *Proceedings of the 24th ACM SIGKDD International Conference on Knowledge Discovery & Data Mining (KDD '18)*. Association for Computing Machinery, New York, NY, USA, 1425–1434.

[6] Nixie S Lesmana, Xuan Zhang, and Xiaohui Bei. 2019. Balancing efficiency and fairness in on-demand ridesourcing. *Advances in Neural Information Processing Systems* 32 (2019).

[7] Yafei Li, Huiling Li, Xin Huang, Jianliang Xu, Yu Han, and Mingliang Xu. 2022. Utility-Aware Dynamic Ridesharing in Spatial Crowdsourcing. *IEEE Transactions on Mobile Computing* (2022), 1–14.

[8] Yafei Li, Huiling Li, Baolong Mei, Xin Huang, Jianliang Xu, and Mingliang Xu. 2023. Fairness-Guaranteed Task Assignment for Crowdsourced Mobility Services. *IEEE Transactions on Mobile Computing* (2023).

[9] Yafei Li, Ji Wan, Rui Chen, Jianliang Xu, Xiaoyi Fu, Hongyan Gu, Pei Lv, and Mingliang Xu. 2019. Top-$k$ k Vehicle Matching in Social Ridesharing: A Price-Aware Approach. *IEEE Transactions on Knowledge and Data Engineering* 33, 3 (2019), 1251–1263.

[10] Aqsa Ashraf Makhdomi and Iqra Altaf Gillani. 2023. GNN-based passenger request prediction. *Transportation Letters* (2023). https://doi.org/10.1080/19427867.2023.2283949

[11] Aqsa Ashraf Makhdomi and Iqra Altaf Gillani. 2023. A greedy approach for increased vehicle utilization in ridesharing networks. *arXiv preprint arXiv:2304.01225* (2023).

[12] Aqsa Ashraf Makhdomi and Iqra Altaf Gillani. 2023. Towards a Greener and Fairer Transportation System: A Survey of Route Recommendation Techniques. *ACM Trans. Intell. Syst. Technol.* (2023).

[13] The News Minute. 2021. Ola, Uber score lowest on fairness scale report, Flipkart highest. https://www.thenewsminute.com/article/ola-uber-score-lowest-fairness-scale-report-flipkart-highest-159303. Accessed: 2022-02-24.

[14] Masato Ota, Yuko Sakurai, Mingyu Guo, and Itsuki Noda. 2022. Mitigating Fairness and Efficiency Tradeoff in Vehicle-Dispatch Problems. In *International Conference on Practical Applications of Agents and Multi-Agent Systems*. Springer, 307–319.

[15] Shiyou Qian, Jian Cao, Frédéric Le Mouël, Issam Sahel, and Minglu Li. 2015. SCRAM: A Sharing Considered Route Assignment Mechanism for Fair Taxi Route Recommendations. In *Proceedings of the 21th ACM SIGKDD International Conference on Knowledge Discovery and Data Mining (KDD '15)*. Association for Computing Machinery, New York, NY, USA, 955–964.

[16] Boting Qu, Wenxin Yang, Ge Cui, and Xin Wang. 2020. Profitable Taxi Travel Route Recommendation Based on Big Taxi Trajectory Data. *IEEE Transactions on Intelligent Transportation Systems* 21, 2 (2020), 653–668.

[17] Huigui Rong, Qun Zhang, Xun Zhou, Hongbo Jiang, Da Cao, and Keqin Li. 2020. Tesla: A centralized taxi dispatching approach to optimizing revenue efficiency with global fairness.

[18] Maximilian Schreieck, Hazem Safetli, Sajjad Ali Siddiqui, Christoph Pflügler, Manuel Wiesche, and Helmut Krcmar. 2016. A Matching Algorithm for Dynamic Ridesharing. *Transportation Research Procedia* 19 (2016), 272–285.

[19] Robert Sedgewick and Kevin Wayne. 2011. Algorithms (4th edn). *Addison Wesley Professional* (2011).

[20] Dingyuan Shi, Yongxin Tong, Zimu Zhou, Bingchen Song, Weifeng Lv, and Qiang Yang. 2021. Learning to Assign: Towards Fair Task Assignment in Large-Scale Ride Hailing. In *Proceedings of the 27th ACM SIGKDD Conference on Knowledge Discovery & Data Mining (KDD '21)*. Association for Computing Machinery, New York, NY, USA, 3549–3557.

[21] Business Standard. 2022. Uber brings back carpooling service under new name 'UberX Share' in US. https://www.business-standard.com/article/international/uber-brings-back-carpooling-service-under-new-name-uberx-share-in-us-122062200650_1.html. Accessed: 2023-04-07.

[22] Jiahui Sun, Haiming Jin, Zhaoxing Yang, Lu Su, and Xinbing Wang. 2022. Optimizing Long-Term Efficiency and Fairness in Ride-Hailing via Joint Order Dispatching and Driver Repositioning. In *Proceedings of the 28th ACM SIGKDD Conference on Knowledge Discovery and Data Mining*. 3950–3960.

[23] Na Ta, Guoliang Li, Tianyu Zhao, Jianhua Feng, Hanchao Ma, and Zhiguo Gong. 2018. An Efficient Ride-Sharing Framework for Maximizing Shared Route. *IEEE Transactions on Knowledge and Data Engineering* 30, 2 (2018), 219–233.

[24] Yongxin Tong, Libin Wang, Zhou Zimu, Bolin Ding, Lei Chen, Jieping Ye, and Ke Xu. 2017. Flexible online task assignment in real-time spatial data. *Proceedings of the VLDB Endowment* 10, 11 (2017), 1334–1345.

[25] Yongxin Tong, Yuxiang Zeng, Zimu Zhou, Lei Chen, and Ke Xu. 2022. Unified Route Planning for Shared Mobility: An Insertion-Based Framework. *ACM Trans. Database Syst.* 47, 1, Article 2 (2022), 48 pages.

[26] Jiachuan Wang, Peng Cheng, Libin Zheng, Chao Feng, Lei Chen, Xuemin Lin, and Zheng Wang. 2020. Demand-Aware Route Planning for Shared Mobility Services. *Proc. VLDB Endow.* 13, 7 (2020), 979–991.

[27] Benwei Wu, Kai Han, and Enpei Zhang. 2023. On the task assignment with group fairness for spatial crowdsourcing. *Information Processing & Management* 60, 2 (2023), 103175.

[28] Chak Fai Yuen, Abhishek Pratap Singh, Sagar Goyal, Sayan Ranu, and Amitabha Bagchi. 2019. Beyond Shortest Paths: Route Recommendations for Ride-Sharing. In *The World Wide Web Conference (WWW '19)*. Association for Computing Machinery, New York, NY, USA, 2258–2269.

[29] Yan Zhao, Kai Zheng, Ziwei Wang, Liwei Deng, Bin Yang, Torben Bach Pedersen, Christian S Jensen, and Xiaofang Zhou. 2023. Coalition-based task assignment with priority-aware fairness in spatial crowdsourcing. *The VLDB Journal* (2023), 1–22.

[30] Stephen M Zoepf, Stella Chen, Paa Adu, and Gonzalo Pozo. 2018. The economics of ride-hailing: Driver revenue, expenses and taxes. *CEEPR WP* 5, 2018 (2018), 1–38.